\documentclass{article}
\usepackage{amsmath}
\usepackage{amsfonts}
\usepackage{amssymb}
\usepackage{graphicx}
\font\Sets=msbm10
\def\Integer {\hbox{\Sets Z}}

\begin{document}

\title{Self-similar solutions of NLS-type dynamical systems\footnote{Work supported in
part by PRIN 97 ``Sintesi''.}}
\author{M. Boiti$^{1}$, V.G. Marikhin$^{2}$,
F. Pempinelli$^{1}$, and A.B. Shabat$^{2}$\\ \\
$^{1}$Dipartimento di Fisica dell'Universit\`{a} and Sezione INFN, Lecce, Italy
\\
$^{2}$L.D. Landau Institute for Theoretical Physics, Moscow, Russia}
\maketitle
\begin{abstract}
We study self-similar solutions of NLS-type dynamical systems.
Lagrangian
approach is used to show that they can be reduced to three canonical forms,
which are related by Miura transformations. The fourth Painlev\'{e} equation
(PIV) is central in our consideration - it connects Heisenberg model,
Volterra model and Toda model to each other.
The connection between the rational solutions
of PIV and Coulomb gas in a parabolic potential is established.
We discuss also the possibility to obtain an
exact solution for optical soliton i.e. of the NLS equation with
time-dependent dispersion.
\end{abstract}

\section{Introduction}

The self-similar solutions are a very useful instrument to investigate some
important physical applications of integrable systems.

There are many examples of these applications.

The importance of this consideration follows from the fact that the asymptotic
form of any solution is a self-similar one. Because we deal with integrable
systems we can build exact self-similar solutions from their asymptotic. This
approach looks like a form of Painlev\'{e} test. In this article we establish
a connection between the self-similar solutions of the NLS-type dynamical
systems and the fourth Painlev\'{e} equation (PIV).

The purpose of this article is to obtain the equations for the self-similar
solutions of NLS-type equations. We show that any member of this NLS family
can be reduced by translation and dilation transformations to three
fundamental canonical forms. Miura transformations among these canonical forms
and B\"{a}cklund transformations will be used to construct new solutions. We
show that the Miura transformations are preserved also under the self-similar
reduction.

We find also an example of an exact self-similar solution for the NLS equation
with time-dependent dispersion.

The structure of the article is the following. In the next section we consider
the Hamiltonian formalism for a NLS-like family of equations and we present
their possible canonical forms and Miura transformations relating them. In
Section 3 we derive the equations for the corresponding self-similar solutions
and show that all possible cases are connected with the Painlev\'{e} IV
equation (PIV). We present a few exact solutions, interesting for some
physical applications, namely a Toda system corresponding to a ``collapsed''
solution of the NLS equations and the Heisenberg system describing spin dynamics.

\section{Canonical forms and Miura transformations}

Hamiltonians of NLS-type dynamical systems have the form:
\begin{equation}
H=P_{X}Q_{X}+\epsilon P^{2}Q_{X}^{2}+\alpha PQ_{X}^{2}+\beta Q_{X}^{2}+\gamma
P^{2}Q_{X}+\delta P^{2},\label{ham}
\end{equation}
where $\epsilon,\,\alpha,\,\beta,\,\gamma\,$and $\delta$ are arbitrary parameters.

By using translations
\begin{equation}
P\rightarrow p-p_{0},\quad Q\rightarrow q-x\,q_{0x}-tq_{0t},\quad X\rightarrow
x-v\,t,\label{spq}
\end{equation}
and dilations of the canonical variables $P,\;Q$ and space $X$ and by
re-scaling the Hamiltonian, the Hamiltonian (\ref{ham}) can be transformed
into one of the following canonical forms.

\subsection{Canonical forms}

\begin{description}
\item [I.]\textbf{Heisenberg model} $\epsilon\neq0\quad$ Hamiltonian
(\ref{ham}) by means of the transformation
\[
P=p-\frac{\alpha}{2\epsilon},\quad Q=q-\frac{\gamma}{2\epsilon}x-\frac
{\alpha\delta}{\epsilon}t,\quad X=x+\frac{\alpha\gamma}{\epsilon}t
\]
reduces to
\begin{equation}
H=p_{x}q_{x}+\epsilon p^{2}q_{x}^{2}+(\delta-\frac{\gamma^{2}}{4\epsilon
})p^{2}+(\beta-\frac{\alpha^{2}}{4\epsilon})q_{x}^{2}.\label{hamh}
\end{equation}
By the dilations $p\rightarrow\displaystyle-\frac{p}{\epsilon}$ and
$H\rightarrow\displaystyle-\frac{1}{\epsilon}H$ and by renaming the constants
we have
\begin{equation}
H=p_{x}q_{x}-p^{2}q_{x}^{2}+\delta p^{2}+\beta q_{x}^{2}.\label{hhgeneral}
\end{equation}
The generic case with $\delta\neq0$ and $\beta\neq0$ is considered in Section
3.3. Of special interest are the two special following cases.

\begin{description}
\item [A]Isotropic Heisenberg model $(\beta=0,\;\delta=0)$, i.e.
\begin{equation}
H=p_{x}q_{x}-p^{2}q_{x}^{2}.\label{hh}
\end{equation}

\item[B] Anisotropic Heisenberg model $(\beta\neq0,$ $\delta=0)$, i.e.
\begin{equation}
H=p_{x}q_{x}-p^{2}q_{x}^{2}+\beta q_{x}^{2}.\label{ha}
\end{equation}

The other case $(\delta\neq0,$ $\beta=0)$ is obtained from this one by the
transformation $p\leftrightarrow q_{x},\;\beta\rightarrow\delta.$
\end{description}

\item[II.] \textbf{Volterra model} $(\epsilon=0,\;\alpha\neq0,\;\gamma\neq0.)$
Hamiltonian (\ref{ham}) reduces to
\begin{equation}
H=p_{x}q_{x}+p\,q_{x}^{2}+p^{2}q_{x}.\label{hv}
\end{equation}
by the transformation
\begin{equation}
P=p-\frac{\beta}{\alpha},\quad Q=q-\frac{\delta}{\gamma}x-\frac{\delta
^{2}\alpha}{\gamma^{2}}t-\frac{2\beta\delta}{\alpha}t,\quad\,X=x+2\left(
\frac{\alpha\delta}{\gamma}+\frac{\gamma\beta}{\alpha}\right)  t
\end{equation}
and by the dilation
\[
p\rightarrow\frac{p}{\gamma},\quad q\rightarrow\frac{q}{\alpha}.
\]

\item[III.] \textbf{Toda model} $(\epsilon=0,\;\alpha\neq0,\;\gamma=0.)$
Hamiltonian (\ref{ham}) reduces to
\begin{equation}
H=p_{x}q_{x}+p\,q_{x}^{2}+p^{2}.\label{ht}
\end{equation}
by the sequence of transformations
\begin{align}
P  & =p-\frac{\beta}{\alpha},\quad Q=q-2\frac{\beta\delta}{\alpha}t\\
\quad p  & \rightarrow\frac{p}{\delta},\quad q\rightarrow\frac{q}{\alpha}.
\end{align}
\end{description}

Note that $p\leftrightarrow q_{x}\iff\alpha\leftrightarrow\gamma
,\;\beta\leftrightarrow\delta\;.$ Other cases lead to linear problems.

\subsection{Miura transformations}

One can easily verify that the NLS-type systems are connected by Miura
transformations according to the following scheme:
\begin{equation}
\text{isotropic Heisenberg\thinspace model\thinspace(I)}\rightarrow
\text{Volterra\thinspace model\thinspace(II)}\rightarrow\text{Toda\thinspace
model\thinspace(III)}\label{mi}
\end{equation}
\[
\begin{array}
[c]{lll}
\text{I}\rightarrow\text{II}: & \displaystyle-p\,q_{x}+\frac{p_{x}}
{p}\rightarrow p, & -p\,q_{x}\rightarrow q_{x}\\
\text{II}\rightarrow\text{III}: & p\,q_{x}+p_{x}\rightarrow p, &
p+q_{x}\rightarrow q_{x}.
\end{array}
\]

\section{Self-similar solutions}

Let us choose solutions for the NLS-type systems in the form
\begin{equation}
q_{x}=b(t)\,y(\xi),\quad p=c(t)\,z(\xi),\quad\xi=x\,a(t),\label{sam}
\end{equation}

To obtain equations for $y$ and $z$ one can consider $p$ as a lagrangian
coordinate, $\xi$ as the new time, substitute ansatz (\ref{sam}) into the
Lagrangian
\begin{equation}
L_{1}=p\,q_{t}-H,\label{lag1}
\end{equation}
and require that the dependence on $t$ factorizes (i.e. the dependence on $t$
of all terms in Lagrangian must be the same). Then, by considering the
variation of the corresponding action with respect to $z$ one gets an equation
for $y$ and $z.$

In the same way one considers $q$ as a lagrangian coordinate and requires that
the $t$ dependence in the Lagrangian
\begin{equation}
L_{2}=-qp_{t}-H,\label{lag2}
\end{equation}
factorizes. Finally, evaluating the variation of the action with respect to
$y$ one obtains the second equation for $y$ and $z.$

Lagrangians (\ref{lag1}) and (\ref{lag2}) can be rewritten as
\[
L_{i}=L_{0i}-V,
\]
where the ``bare'' parts of the Lagrangians
\begin{equation}
L_{01}=p\,q_{t}-p_{x}q_{x},\quad L_{02}=-q\,p_{t}-p_{x}q_{x}\label{h0}
\end{equation}
are the same in all models. Requiring the factorization of time for both
$L_{oi}$ one gets
\begin{equation}
a^{\prime}=2a^{3},\quad b=a^{\nu},\quad c=a^{\lambda},\label{uh0}
\end{equation}
and then the ``bare'' parts of the actions have the form
\begin{align}
S_{01} &  =\int dt\,b(t)\,c(t)\int d\xi\lbrack2(\nu-1)zY+2\xi zy-z^{\prime
}y],\quad Y^{\prime}=y\label{s01}\\
S_{02} &  =\int dt\,b(t)\,c(t)\int d\xi\lbrack-2\lambda zY-2\xi z^{\prime
}Y-z^{\prime}y].\label{s02}
\end{align}
By variation on $z$ of the first action and on $Y$ of the second action one
derives the system of equations
\begin{equation}
\begin{array}
[c]{l}
y^{\prime}+2\xi y+2(\nu-1)Y+c_{1}=\frac{\partial v(y,z)}{\partial z}\\
-z^{\prime}+2\xi z+2(\lambda-1)Z+c_{2}=\frac{\partial v(y,z)}{\partial
y},\quad\quad Z^{\prime}=z
\end{array}
\label{sys}
\end{equation}
for $\nu$ and $\lambda$ such that $V(p,q_{x})=V(cz,by)=abc\,v(z,y).$ Note
that, respectively, for $\nu\neq1$ and for $\lambda\neq1$ the constant $c_{1}$
and $c_{2}$ can be omitted. Note also that the system is lagrangian only for
$\lambda+\nu=2,$ since only in this case the two actions in (\ref{s01}) and
(\ref{s02}) differ by a total derivative with respect to $\xi$. The next
considerations depend on the model (i.e. on $V$).

\section{Volterra model}

In this case
\begin{equation}
V=pq_{x}^{2}+p^{2}q_{x}\Rightarrow\nu=\lambda=1\label{vo}
\end{equation}
and we obtain the Lagrangian system $(L=y^{\prime}z+2\xi yz-y^{2}
z-z^{2}y+c_{1}z+c_{2}y)$
\begin{equation}
\begin{array}
[c]{l}
y^{\prime}+2\xi y+c_{1}=y^{2}+2zy\\
-z^{\prime}+2\xi z+c_{2}=2yz+z^{2}
\end{array}
\label{syv}
\end{equation}
which is an equivalent form of the PIV ODE \cite{Gromak}. In fact inserting
$z$ in $L$ from the first equation of (\ref{syv}) and writing
\begin{equation}
w(x,a,b)=y(-\xi,c_{1},c_{2}),\quad a=\frac{1}{2}c_{1}-c_{2}+1,\quad
b=-\frac{1}{2}c_{1}^{2}\label{obv}
\end{equation}
we obtain the PIV Lagrangian
\begin{equation}
L=\frac{w_{x}^{2}}{w}+w^{3}+4w^{2}x+4w(x^{2}-a)-\frac{2b}{w}\label{p4v}
\end{equation}

\subsection{Coulomb gas. Rational solutions of PIV.}
Another useful representation for PIV is
\cite{VeSh}
\begin{equation}\label{g3}
    g'_1 = g_1(g_3-g_2)+\alpha_1, \quad
    g'_2 = g_2(g_1-g_3)+\alpha_2, \quad
    g'_3 = g_3(g_2-g_1)+\alpha_3,
\end{equation}
with additional first integral
\begin{equation}\label{xgamma}
 g_1+g_2+g_3=x\gamma,\quad \gamma=\alpha_1+\alpha_2+\alpha_3.
\end{equation}
It easy to see that all of $g_j =w ,\quad j=1,2,3$ are solutions of PIV if
$$\gamma=-2,\quad a=-\frac{c_j}{2},\quad b=-\frac{\alpha_j^2}{2},\quad
c_j =\alpha_{j+2}-\alpha_{j+1}$$

It is well known (see \cite{Ad} for example) that system (\ref{g3})
has the rational solutions if
\begin{equation}\label{rat}
   \alpha_j= {\gamma\over3}\gamma_j, \quad
   \gamma_j\in\Integer, \quad \gamma_1+\gamma_2+\gamma_3=3, \quad
   \gamma_1 \equiv \gamma_2 \equiv \gamma_3\ (\mod 3).
\end{equation}
We write
\begin{equation}
g_j=\frac{1}{3}\gamma x+\partial_x\log(Q_{j+1}/Q_{j-1})
\end{equation}
where $Q_j$ - polynomials and
\begin{equation}\label{12}
(Q_j''-xQ_j'+m_jQ_j)Q_{j+1}+(Q_{j+1}''+xQ_{j+1}'-m_{j+1}Q_{j+1})Q_j=
2Q_j'Q_{j+1}',\quad (\gamma=3).
\end{equation}

For example
\begin{equation}\label{primer}
 \vec \alpha = (-1, 2, 2), \quad\vec Q = (x, 1, 1), \qquad
 \vec \alpha = (-1,5,-1),\quad  \vec Q = (x^2-1, 1, x^2+1),
\end{equation}
\begin{eqnarray*}
 \vec \alpha = ( 1,4,-2), \quad \vec Q = (x, 1, x^2+1), \qquad
 \vec \alpha = (5,-4,2),\quad  \vec Q = (x,x^4+2x^2-1, x^2+1).
\end{eqnarray*}

It is important for us that any divergence of PIV has a form
\begin{equation}\label{ask4}
w\sim \frac{c}{\xi-\xi_0}-\xi_0 ,\,c^2=1.
\end{equation}
It is well known also that the asymptotical behavior of functions
$g_i$ defines by
\begin{equation}\label{cen}
g_i \rightarrow -\frac{2}{3}\xi, \quad \xi\rightarrow\infty,
\end{equation}
либо
\begin{equation}\label{ugl}
g_1 \rightarrow -2\xi, \,\,g_2,g_3 \rightarrow 0
 \quad \xi\rightarrow\infty,
\end{equation}

We are looking for rational solutions of the PIV equation in a
form
\begin{equation}\label{qul}
w=-\delta\xi+\sum\limits_i \frac{c_i}{\xi-\xi_i},
\end{equation}
where $\delta=\frac{2}{3}$ in a case (\ref{cen}) and
$\delta=0$ or $\delta=2$ in (\ref{ugl}).

Substituting (\ref{qul}) to (\ref{ask4}) one has
\begin{equation}\label{qgaz}
(\delta-1)\xi_i =\sum\limits_{j\neq i}
\frac{c_j}{\xi_i -\xi_j},\quad c_i =\pm 1.
\end{equation}
Equation (\ref{qgaz}) describes the gas of charged particles
with Coulomb interaction in a parabolic potential, and the level
statistics in a hermitian case because one can obtain
(\ref{qgaz}) by variation of a functional
\begin{equation}\label{Uf}
U=\sum\limits_i \frac{1}{2}(1-\delta)c_i \xi_i^2 +\sum\limits_{i\neq j}
c_i c_j\log(\xi_i -\xi_j ).
\end{equation}

In the case when all charges equal $c_i =1,$ (or $c_i =-1,)$
one have a famous solution of PIV
\begin{equation}\label{rp4}
w=-2\xi+\frac{H'_n (\xi)}{H_n (\xi)},
\end{equation}
where $H_n (\xi)- $ Hermitian polynomials (cр.,\cite{EC}).
So, the roots of $H_n$
coincide with distribution of the positive (negative)
charges  in the potential (\ref{Uf}).

For example, solution $\vec{Q}(1,4,2)$ (see (\ref{primer})) defines
three solutions of (\ref{qgaz}) if $\delta=\frac{2}{3},$
two are "trivial" (Hermitian polynomials)
and third describes a system of two positive and one negative charges:
\begin{equation}\label{dde}
\xi_{1,2}=\pm i,\quad \xi_3 =0,\quad c_{1,2}=1,\quad c_3 =-1.
\end{equation}
In general case one have three solutions of (\ref{qgaz}):
all positive charges coincide with the roots of $Q_{j+1},$
all negative charges coincide with the roots of $Q_{j-1};$
numbers of charges are $N_{\pm}=deg\,Q_{j\pm 1}$.

\section{Heisenberg model}

In this case
\begin{equation}
V=-p^{2}q_{x}^{2}\Rightarrow\lambda=1-\nu,\label{he}
\end{equation}
and from (\ref{sys}) we have
\begin{equation}
\begin{array}
[c]{l}
y^{\prime}+2\xi y+2(\nu-1)Y=-2zy^{2}\\
-z^{\prime}+2\xi z-2\nu Z=-2z^{2}y.
\end{array}
\label{syh}
\end{equation}
This system admits the conservation law
\begin{equation}
zy^{\prime}-yz^{\prime}+2\xi zy+3z^{2}y^{2}=\mu\label{mu}
\end{equation}
where $\mu$ is an arbitrary constant. Inserting $z$ from the first equation of
(\ref{syh}) into this conservation law we get a third order ODE satisfied by
$Y$
\begin{equation}
\frac{y^{\prime\prime}}{y}-\frac{3}{2}\frac{y^{\prime2}}{y^{2}}+8\left(
\nu-1\right)  \xi\frac{Y}{y}+6\left(  \nu-1\right)  ^{2}\frac{Y^{2}}{y^{2}
}+2\xi^{2}+2\left(  \nu-\mu\right)  =0.
\end{equation}
By setting
\begin{equation}
Y=e^{\int^{\xi}d\xi^{\prime}w(\xi^{\prime})}.\label{Yw}
\end{equation}
we have
\begin{equation}
\frac{w^{\prime\prime}}{w}-\frac{3}{2}\frac{w^{\prime2}}{w^{2}}-\frac{1}
{2}w^{2}+8\xi(\nu-1)\frac{1}{w}+6(\nu-1)^{2}\frac{1}{w^{2}}+2\xi^{2}+2\left(
\nu-\mu\right)  =0.
\end{equation}
Finally, by means of the transformation
\begin{equation}
w=\frac{2(\nu-1)}{f}\label{wf}
\end{equation}
one gets the PIV ODE
\begin{equation}
f^{\prime\prime}=\frac{1}{2}\frac{f^{\prime2}}{f}+\frac{3}{2}f^{3}+4\xi
f^{2}+2(\xi^{2}+\nu-\mu)f-2\frac{(\nu-1)^{2}}{f}.\label{fPIV}
\end{equation}

\subsection{Self-similar solutions for spin dynamic}

If we introduce the vector $\overrightarrow{S}$
\begin{align}
S_{1} &  =p(q^{2}-1)+q\\
S_{2} &  =ip(q^{2}+1)+iq\\
S_{3} &  =2pq+1
\end{align}
of unit length, i.e. $\overrightarrow{S}^{2}=1$, the Hamiltonian (\ref{hh})
for the isotropic Heisenberg model can be rewritten as
\begin{equation}
H=-\frac{1}{4}\overrightarrow{S}_{x}^{2}
\end{equation}
and the corresponding evolution equations
\begin{align}
p_{t} &  =\left(  p_{x}-2p^{2}q_{x}\right)  _{x}\\
q_{t} &  =-q_{xx}-2q_{x}^{2}p.
\end{align}
as
\begin{equation}
i\overrightarrow{S}_{t}=\overrightarrow{S}\times\overrightarrow{S}_{xx}
\end{equation}
which up to a change to the physical time $t\rightarrow it$ is the system of
equations describing the Heisenberg Ferromagnetic model.

In the previous section we show that the general self-similar solution is
related to the PIV Painlev\'{e} transcendents. It is interesting to note that
the requirement
\begin{equation}
S_{3}=\sigma,\quad\quad\sigma=\pm1
\end{equation}
is compatible with the existence of self-similar solutions. More precisely, the
solutions are related to the parabolic cylindric functions. In fact from
\begin{equation}
S_{3}=1+2zY
\end{equation}
inserting $z$ from the first equation of the system (\ref{syh}) and using
(\ref{Yw}) and (\ref{wf}) we have
\begin{equation}
f^{\prime}=2\sigma(\nu-1)+2\xi f+f^{2}\label{fsigma}
\end{equation}
and one can verify directly that these $f$ are solutions of the PIV equation
(\ref{fPIV}) for
\begin{equation}
\nu-\mu=1+\sigma(\nu-1).
\end{equation}
The Riccati equation (\ref{fsigma}) can be linearized by writing
\begin{equation}
f=-\xi-\frac{\psi^{\prime}}{\psi}.
\end{equation}
The $\psi$ satisfy the Weber equation
\begin{equation}
\psi^{\prime\prime}-\left(  \xi^{2}-2\sigma(\nu-1)-1\right)  \psi=0,
\end{equation}
whose general solution is a linear combination of the parabolic cylindric
functions
\begin{align}
\psi_{1} &  =D_{\sigma(\nu-1)}(\sqrt{2}\xi)\\
\psi_{2} &  =D_{\sigma(\nu-1)}(-\sqrt{2}\xi).
\end{align}

\subsection{Deformation of Heisenberg model}

First, let us note that the evolution equations corresponding to the general
Hamiltonian (\ref{hhgeneral}) with $\beta\neq0$ and $\delta\neq0$ do not admit
self-similar solutions of the form (\ref{sam}). In the case
\begin{equation}
V=-p^{2}q_{x}^{2}+\beta q_{x}^{2}
\end{equation}
the evolution equations are
\begin{align}
p_{t} &  =(p_{x}-2p^{2}q_{x}+2\beta q_{x})_{x}\label{epsilonnotzero1}\\
q_{t} &  =-q_{xx}-2q_{x}^{2}p\label{epsilonnotzero2}
\end{align}
and self-similar solutions are obtained for $\nu=1$, $\,\lambda=0.$

We have
\begin{align}
2z &  =g^{\prime}-2\xi g+c_{1}g^{2}\\
y &  =\frac{1}{g}\quad\quad
\end{align}
and
\begin{equation}
g^{\prime\prime\prime}=4\xi g+3c_{1}^{2}g^{2}g^{\prime}-4c_{1}\xi gg^{\prime
}-4c_{1}g^{2}+\left(  \frac{g^{\prime2}-4\beta}{g}\right)  ^{\prime
}\label{y3similar}
\end{equation}
with $c_{1}$ a constant of integration. By introducing $g=e^{w}$ we have
\begin{align}
w^{\prime\prime\prime} &  =-w^{\prime}w^{\prime\prime}+4\xi+4\beta w^{\prime
}e^{-2w}+\nonumber\\
&  +3c_{1}^{2}w^{\prime}e^{2w}-4c_{1}\xi w^{\prime}e^{w}-4c_{1}e^{w}
\end{align}
that can be easily integrated once to
\begin{align}
w^{\prime\prime} &  =-\frac{1}{2}w^{\prime2}+2\xi^{2}-2\beta e^{-2w}
+\frac{\gamma}{2}+\nonumber\\
&  +\frac{3}{2}c_{1}^{2}e^{2w}-4c_{1}\xi e^{w}.
\end{align}
where $\gamma$ is a constant of integration. Coming back to $g$ we get
\begin{align}
g^{\prime\prime} &  =\frac{1}{2}\frac{g^{\prime2}}{g}+2\left(  \xi^{2}
+\gamma\right)  g-2\frac{\beta}{g}+\nonumber\\
&  +\frac{3}{2}c_{1}^{2}g^{3}-4c_{1}\xi g^{2}.\label{PIVc}
\end{align}
For $c_{1}\neq0$ changing $c_{1}g\rightarrow g$ we get the PIV ODE equation.
Dividing (\ref{PIVc}) by $g$, differentiating with respect to $\xi$ and then
by inserting in the obtained equation $g^{\prime\prime}$ from (\ref{PIVc}) one
gets
\begin{equation}
g^{\prime\prime\prime}=4\xi g+4\left(  \xi^{2}+\gamma\right)  g^{\prime
}+6c_{1}^{2}g^{2}g^{\prime}-12c_{1}\xi gg^{\prime}-4c_{1}g^{2}\label{y3}
\end{equation}
which for $c_{1}=1,$ together with (\ref{y3similar}), can be considered an
equivalent form of the PIV equation.

Of special interest is the case $c_{1}=0$, i.e.
\begin{equation}
g^{\prime\prime\prime}=4\xi g+4\left(  \xi^{2}+\gamma\right)  g^{\prime}.
\end{equation}
One can easily verify that its general solution is given by
\begin{equation}
g=d_{1}\psi_{1}^{2}+d_{2}\psi_{2}^{2}+d_{3}\psi_{1}\psi_{2}\label{c1c2c3}
\end{equation}
where $d_{j}$ ($j=1,2,3$) are arbitrary constants and $\psi_{1},$ $\psi_{2} $
are two independent solutions of the Weber equation
\begin{equation}
\psi^{\prime\prime}-\left(  \xi^{2}+\gamma\right)  \psi=0,
\end{equation}
e.g. the parabolic cylindric functions
\begin{align}
\psi_{1} &  =D_{-\frac{\gamma+1}{2}}(\sqrt{2}\xi)\\
\psi_{2} &  =D_{-\frac{\gamma+1}{2}}(-\sqrt{2}\xi).
\end{align}

\subsection{Landau-Lifshitz model.}

The Hamiltonian of the Landau-Lifshitz model
\begin{equation}
H=p_{x}q_{x}-p^{2}(q_{x}^{2}+r(q))-p\frac{r^{\prime}(q)}{2}-\frac
{r^{\prime\prime}(q)}{12},\quad r^{V}=0\label{hll}
\end{equation}
corresponds to the ``diffusion'' spin dynamics
\begin{equation}
i\vec{S}_{t}=[\vec{S}\times(\vec{S}_{xx}+\hat{J}\vec{S})],\quad\vec{S}
^{2}=1,\label{sll}
\end{equation}
where we introduced the ``canonical'' parametrization of the vector $\vec{S}$
\begin{equation}
\vec{S}=(S_{1},S_{2},S_{2})=(p(q^{2}-1)+q,\,ip(q^{2}+1)+iq,\,2pq+1),\quad
r(q)=-\frac{1}{4}(\vec{S}_{p}\hat{J}\vec{S}_{p}).\label{spar}
\end{equation}

One can obtain another parametrization of $\vec{S}$ from a
spectral problem of the Toda model (see (\ref{zct})):
\begin{equation}\label{spsi}
  w(x)S_3=(\psi_1\psi_2)_x,\quad w(x)(S_1+iS_2)=i(\psi_2^2)_x,
                           \quad w(x)(S_1-iS_2)=i(\psi_1^2)_x.
\end{equation}
where $w=\psi_{1,x}\psi_2-\psi_{2,x}\psi_1 .$

It is important that the Frenet equations
\begin{equation}\label{fre}
  \vec{S}_x=k\vec{N},\quad
 \vec{N}_x=-k\vec{S}+\chi\vec{B},\quad \vec{B}_x=-\chi\vec{N},
\quad (\vec{S},\vec{S})=1,
\end{equation}
 where $\vec{B}=\vec{S}\times\vec{N}-$ binormal, can be obtained
 from (\ref{zct}) if
\begin{equation}\label{ct}
k^2=4p,\quad i\chi+ q_x + (\log k)_x=\lambda,
\end{equation}

\section{Toda model}

In this case
\begin{equation}
V=pq_{x}^{2}+p^{2}\Rightarrow\lambda=2,\quad\nu=1,\label{te}
\end{equation}
From (\ref{sys}) we have
\begin{equation}
\begin{array}
[c]{c}
y^{\prime}+2\xi y+C=y^{2}+2z\\
-z^{\prime}+2\xi z+2Z=2zy.
\end{array}
\label{stoda}
\end{equation}
Excluding $y$ one obtains the third-order equation on $Z\,$
\begin{equation}
Z^{\prime\prime\prime}=\frac{Z^{\prime\prime2}}{2Z^{\prime}}+2(Z^{\prime}
\xi^{2}-\frac{Z^{2}}{Z^{\prime}})-4Z^{\prime2}+2(C+2)Z^{\prime}\label{ovc}
\end{equation}
which is integrable to
\begin{equation}
Z^{\prime\prime2}+4Z^{\prime3}-4(C+2)Z^{\prime2}+4DZ^{\prime}=4(\xi Z^{\prime
}-Z)^{2},\quad D=c_{2}(c_{1}+2).\label{shab}
\end{equation}

Zero-curvature representation for Toda model (\ref{ht}) has a form
\begin{equation}\label{zct}
\psi_{xx}+(q_x -\lambda)\psi_x +p\psi=0,\quad
\psi_t =(q_x -\lambda)\psi_x -p\psi
\end{equation}
One can substitute self-similar form (\ref{stoda}) to obtain
zero-curvature representation for PIV:
\begin{equation}\label{zct4}
\psi_{\xi\xi}+(y -\lambda)\psi_{\xi} +z\psi=0,\quad
\lambda\psi_{\lambda} =(\lambda+\xi+y)\psi_{\xi} -z\psi ,
\end{equation}
where $w=y+\xi$ is solution of PIV.

As we shall show in the following section, (\ref{shab}) is another equivalent
form of the PIV equation. It is interesting to note that it can be considered
a conservation law of the system (\ref{stoda}). In fact inserting $Z$ from the
second equation of (\ref{stoda}) into (\ref{shab}) we get
\begin{equation}
z^{\prime}(y-2\xi)+z(y-2\xi)^{2}-z^{2}+(C+2)z=D.\label{Todaconservation}
\end{equation}

The function $Z$ can be also expressed in terms of three PIV transcendent
$y_{n}(\xi)$ ($n\in Z_{3}$) according to the formula
\begin{equation}
2Z=y_{1}y_{2}y_{3}-\alpha_{1}y_{2}+\alpha_{2}y_{1}
\end{equation}
where the parameters $\alpha_{i}$ are defined by
\begin{equation}
C=\alpha_{2}-\alpha_{1}-2,\quad D=-\alpha_{1}\alpha_{2}.\label{al}
\end{equation}
and the functions $y_{n}$ are related by the equation
\begin{equation}
y_{1}+y_{2}+y_{3}=2\xi\label{sumy}
\end{equation}
and satisfy the PIV equations
\begin{equation}
y_{n}^{\prime\prime}=\frac{y_{n}^{\prime2}}{2y_{n}}+\frac{3}{2}y_{n}^{3}-4\xi
y_{n}^{2}+2(\xi^{2}+\widetilde{\gamma}_{n})y_{n}+\frac{1}{2}\frac
{\widetilde{\beta}_{n}}{y_{n}}\label{PIVn}
\end{equation}
with
\begin{equation}
\widetilde{\gamma}_{n}=\frac{\alpha_{n+1}-\alpha_{n-1}}{2}
\end{equation}
\begin{equation}
\widetilde{\beta}_{n}=-\alpha_{n}^{2}
\end{equation}
\begin{equation}
\alpha_{1}+\alpha_{2}+\alpha_{3}=-2.
\end{equation}
This result can be obtained using the connection among nonlinear chains and
Painlev\'{e} equations found in \cite{Ad}. In fact let us consider the
nonlinear chain
\begin{equation}
y_{n}^{\prime}+y_{n}(y_{n+1}-y_{n-1})+\alpha_{n}=0,\quad\quad n\in Z_{3}
\end{equation}
where the constants $\alpha_{n}$ are normalized as follows
\begin{equation}
\alpha_{1}+\alpha_{2}+\alpha_{3}=-2
\end{equation}
and the $y_{n}$'s satisfy (\ref{sumy}).

Easy computation shows that indeed the $y_{n}$'s satisfy the PIV equations
(\ref{PIVn}). Let us now introduce
\begin{equation}
h=y_{1}y_{2}
\end{equation}
Then we have
\begin{equation}
Z^{\prime}=h
\end{equation}
and
\begin{equation}
\frac{h^{\prime}}{h} =y_{1}\left(  1-\frac{\alpha_{2}}{h}\right)
-y_{2}\left(  1+\frac{\alpha_{1}}{h}\right)
\end{equation}
\begin{equation}
-\frac{2f}{h}+2\xi=y_{1}\left(  1-\frac{\alpha_{2}}{h}\right)  +y_{2}\left(
1+\frac{\alpha_{1}}{h}\right)  .
\end{equation}
From these equations, eliminating $y_{1}$ and $y_{2},$ we get finally that $f
$ satisfies (\ref{shab}).

The system (\ref{te}) is equivalent (up to the change $t\rightarrow it$) to a
system of two coupled NLS equations. In fact if we rewrite it as
\begin{equation}
p_{t} =(p_{x}+2pq_{x})_{x}
\end{equation}
\begin{equation}
q_{t} =-q_{xx}+q_{x}^{2}+2p
\end{equation}
and introduce
\begin{equation}
u =e^{-q}
\end{equation}
\begin{equation}
p =uv
\end{equation}
we get
\begin{equation}
u_{t}+u_{xx}+2u^{2}v =0
\end{equation}
\begin{equation}
v_{t}-v_{xx}-2v^{2}u =0.
\end{equation}

\section{NLS equation with time-dependent dispersion.}
\subsection{Integrable case}
The special case of the NLS equation deserves a more detailed analysis and an
extension of the study to some nonintegrable variants. Let us consider the NLS
equation with time dependent coefficients
\begin{equation}
iq_{t}+f(t)q_{xx}+g(t)|q|^{2}q=0.
\end{equation}
Renaming the time
\begin{equation}
t\rightarrow\int^{t}f(t^{\prime})dt^{\prime}
\end{equation}
and the function $g(t)$
\begin{equation}
g(t)\rightarrow\frac{g\left(  t\right)  }{f\left(  t\right)  }
\end{equation}
it can be rewritten as
\begin{equation}
iq_{t}+q_{xx}+g(t)\left|  q\right|  ^{2}q=0.
\end{equation}
It admits self-similar solutions for $g(t)$ of the form
\begin{equation}
g(t)=\alpha(4t)^{\frac{m-1}{2}}
\end{equation}
where $m\in R$ and the constant coefficient $\alpha$ can be normalized to
$\pm1,$ the sign being most important. The self-similar solution (up to a
trivial rescaling and $t>0$) is given by
\begin{equation}
q=\left|  q\right|  e^{i\Theta(\eta,t)},\quad\left|  q\right|  ^{2}
=4k(t)^{m+1}Y^{\prime}(\eta),\quad\eta=k(t)x,\quad k(t)=\frac{1}{2t=^{1/2}}
\end{equation}
where
\begin{equation}
\Theta(\eta,t)=\theta(\eta)+\theta_{0}\log t,\quad\quad\theta_{0}\in R
\end{equation}
\begin{equation}
\theta(\eta)=\frac{\eta^{2}}{2}+m\int\frac{Y}{Y^{\prime}}\,d\eta,
\end{equation}
and $Y$ satisfies the ODE
\begin{equation}
2Y^{\prime\prime\prime}Y^{\prime}-Y^{\prime\prime2}+4\eta^{2}Y^{\prime
2}-4m^{2}Y^{2}+16\alpha Y^{\prime3}-16\theta_{0}Y^{\prime2}=0.\label{Ythirdk}
\end{equation}
A Painlev\'{e} analysis of this equations shows that it has no movable
critical poles only for $m^{2}=1$ and we recover equation (\ref{shab}) up to a
trivial rescaling in order to pass from the variable $\xi$ to the physical
variable $\eta.$ In this case the integration factor $Y^{\prime\prime
}/Y^{\prime2}$ allows to integrate and to reduce it to the second order ODE
\begin{equation}
Y_{\eta\eta}^{2}+4\left(  \eta Y_{\eta}-Y\right)  ^{2}+2\alpha Y_{\eta
}(2Y_{\eta}-\alpha\theta_{0})^{2}+2\alpha\mu^{2}Y_{\eta}
=0.\label{algebraicPIV}
\end{equation}
with $\mu^{2}$ a constant of integration. This self-similar solution of the NLS
equation was first obtained in \cite{BP} in the case $\theta_{0}=0.$ Its
properties can be studied following the same lines as in \cite{BP}.

The solution $Y$ can be expressed as
\begin{equation}
-\alpha Y=\frac{1}{2}W\left(  W-\eta\right)  ^{2}+\frac{1}{8W}\left[  W_{\eta
}^{2}-2W_{\eta}-\mu^{2}+1\right]  +\frac{1}{2}\theta_{0}\left(  W-\eta\right)
.\label{Y=PIV}
\end{equation}
where $W$ satisfies
\begin{equation}
WW_{\eta\eta}=\frac{1}{2}W_{\eta}^{2}-6W^{4}+8\eta W^{3}-2(\eta^{2}-\theta
_{0})W^{2}-\frac{1}{2}\left(  \mu-1\right)  ^{2}
\end{equation}
which up to a trivial change of variable is the PIV ODE. Equation
(\ref{Y=PIV}) can be considered an algebraic resolution of (\ref{algebraicPIV}
) which is quadratic in $Y_{\eta\eta}.$

We have also
\begin{equation}
-2\alpha Y_{\eta}=\left(  W-\eta\right)  ^{2}+\frac{\left(  W_{\eta}
+\mu-1\right)  ^{2}}{4W^{2}}.
\end{equation}

Therefore for $\alpha=-1$ and for $\mu$ and $W$ real the condition $Y_{\eta
}\geq0$ is automatically satisfied. On the contrary for $\alpha=1$ and $\mu$
and $W$ real we have $Y_{\eta}\leq0$ and we have not self-similar solutions. Let
us then choose from now on $\alpha=-1.$

We can also express $W$ in terms of $Y$ as follows
\begin{equation}
W=\frac{-\mu Y_{\eta\eta}+\left(  2Y_{\eta}+\theta_{0}\right)  (2Y+\eta
\theta_{0})+\eta\mu^{2}}{\left(  2Y_{\eta}+\theta_{0}\right)  ^{2}+\mu^{2}}.
\end{equation}

Starting from a suggestion by Bureau \cite[pag. 210]{Bureau} one can solve
algebraically (\ref{algebraicPIV}) in an alternative way and, precisely, one
can express $Y$ in terms of another function $\widehat{W}$ as follows
\begin{equation}
4Y=h\eta+\frac{\widehat{W}_{\eta}^{2}}{2\widehat{W}}-2\widehat{W}^{3}
-4\eta\widehat{W}^{2}+2\left(  \eta^{2}-a\right)  \widehat{W}+\frac
{b}{4\widehat{W}}
\end{equation}
with $\widehat{W}$ satisfying the equation
\begin{equation}
\widehat{W}\widehat{W}_{\eta\eta}=\frac{1}{2}\widehat{W}_{\eta}^{2}
-6\widehat{W}^{4}+8\eta\widehat{W}^{3}-2\left(  \eta^{2}-a-\sigma^{2}\right)
\widehat{W}^{2}+\frac{b}{4}.
\end{equation}
where
\[
\sigma^{4}=-1
\]
and $a,$ $b$ and $h$ are defined by
\begin{equation}
4a^{3}+3(3\mu^{2}-\theta_{0}^{2})a+\theta_{0}\left(  9\mu^{2}+\theta_{0}
^{2}\right)  =0
\end{equation}
\begin{equation}
3b=-2a^{2}-2(3\mu^{2}-\theta_{0}^{2})
\end{equation}
\begin{equation}
h=\frac{2}{3}\left(  a-2\theta_{0}\right)  .
\end{equation}
The roots of the algebraic equation for $a$ are
\begin{equation}
a_{1} =-\theta_{0}
\end{equation}
\begin{equation}
a_{2,3} =\frac{1}{2}\theta_{0}\pm\frac{3}{2}i\mu.
\end{equation}
and then
\begin{equation}
b_{1} =-2\mu^{2}
\end{equation}
\begin{equation}
b_{2,3} =\frac{1}{2}\left(  \theta_{0}\mp i\mu\right) ^{2}
\end{equation}
\begin{equation}
h_{1} =-2\theta_{0}
\end{equation}
\begin{equation}
h_{2,3} =-\theta_{0}\pm i\mu.
\end{equation}
$\widehat{W}$ can be expressed in terms of $Y$.

In the case $a_{1}=-\theta_{0}$ one gets
\begin{equation}
\widehat{W}_{1}=\frac{1}{2}\eta-\frac{1}{2}\frac{Y}{Y_{\eta}}-\frac{\sigma
^{2}}{4}\frac{Y_{\eta\eta}}{Y_{\eta}}
\end{equation}
and in the case $a_{2,3}=\frac{1}{2}\theta_{0}\pm\frac{3}{2}i\mu$
\begin{equation}
\widehat{W}_{2,3}=\frac{\eta Y_{\eta}-Y}{2Y_{\eta}+\theta_{0}\pm i\mu}
-\frac{\sigma^{2}}{2}\frac{Y_{\eta\eta}}{2Y_{\eta}+\theta_{0}\pm i\mu}.
\end{equation}
The $\widehat{W}$ are related by the equation
\begin{equation}
\frac{\zeta_{2}W_{2}}{W_{2}-W_{1}}=\frac{\zeta_{3}W_{3}}{W_{3}-W_{1}}
=\frac{\zeta_{3}W_{3}-\zeta_{2}W_{2}}{W_{3}-W_{2}}
\end{equation}
where
\begin{equation}
\zeta_{2,3}=\frac{1}{2}\left(  \theta_{0}\pm i\mu\right)  .
\end{equation}
The B\"{a}cklund transformation for the NLS equation induces a B\"{a}cklund
transformation for its self-similar solution $Y.$ Precisely one gets a new
self-similar solution $\widetilde{Y}$ which satisfies (\ref{algebraicPIV}) with
$\mu\rightarrow2-\mu$ and which is given by
\begin{equation}
\widetilde{Y}=Y+\frac{1}{2}\frac{\left(  \mu-1\right)  \left[  \left(
2Y_{\eta}+\theta_{0}\right)  ^{2}+\mu^{2}\right]  }{\left[  -\mu Y_{\eta\eta
}+\left(  2Y_{\eta}+\theta_{0}\right)  (2Y+\eta\theta_{0})+\eta\mu^{2}\right]
}.
\end{equation}

\subsection{NLS with time-dependent dispersion. A non integrable case}

Let us consider the Lagrangian for NLS with a time-dependent dispersion
\begin{equation}
L=\int\,dt\int\,dx\left[  i\psi^{\ast}\psi_{t}-d(t)\psi_{x}^{\ast}\psi
_{x}+\frac{g}{2}(\psi^{\ast}\psi)^{2}\right]  .\label{opl}
\end{equation}
The canonical transformation $(\psi,\psi^{\ast})\rightarrow(\rho,\phi)$ where
$\psi=\rho^{1/2}\,e^{i\phi}$ gives us
\begin{equation}
L=\int\,dt\int\,dx\left[  -\rho\phi_{t}-d\rho\phi_{x}^{2}-d\frac{\rho_{x}^{2}
}{4\rho}+\frac{g}{2}\rho^{2}\right]  .\label{orl}
\end{equation}
The equations of motion are given by
\begin{align}
& \rho_{t}+2d\partial_{x}(\rho\phi_{x})=0,\label{oner}\\
& \phi_{t}+d\phi_{x}^{2}-\frac{d}{2}\frac{\rho_{xx}}{\rho}+\frac{d}{4}
\frac{\rho_{x}^{2}}{\rho^{2}}-g\rho=0.
\end{align}
The continuity equation (\ref{oner}) is identically satisfied for $\rho$ of
the form
\begin{equation}
\rho=\lambda(t)\,f(\xi),\quad\phi=-\frac{\lambda^{\prime}(t)}{4d\lambda
(t)^{3}}\xi^{2}+\phi_{0}(t),\quad\xi=\lambda(t)x\label{rh}
\end{equation}
with $\lambda,$ $f,$ and $\phi_{0}$ arbitrary functions.

The Lagrangian (\ref{orl}) transforms to
\begin{equation}
L=\int\,\frac{dt}{\lambda(t)}\int\,d\xi\left[  -f\lambda\phi_{0}^{\prime
}(t)-\frac{d\lambda^{3}}{4}\frac{f^{\;\prime2}}{f}+\frac{g}{2}\lambda
^{2}f^{\;2}+\left\{  \frac{1}{\lambda}\left(  \frac{\lambda^{\prime}
}{4d\lambda}\right)  ^{\prime}-d\lambda\left(  \frac{\lambda^{\prime}
}{2d\lambda^{2}}\right)  ^{2}\right\}  f\xi^{2}\right]  .\label{olf}
\end{equation}
The equations of motions are compatible with a dependence of $f$ only on $\xi$
for a time dependent dispersion $d$ and function $\phi_{0}$ satisfying the
requirements
\begin{equation}
\lambda d=\alpha,\quad\frac{\phi_{0}^{\prime}}{\lambda}=-\beta,\quad
d^{\prime\prime}d=d^{\prime2}-4\alpha^{3}\gamma
\end{equation}
with $\alpha$ and $\beta$ constants. Then the equation of motion for $f$ is
given by
\begin{equation}
\frac{\alpha}{2}\left[  \frac{f^{\prime\prime}}{f}-\frac{1}{2}\frac
{f^{\prime2}}{f^{2}}\right]  +\gamma\xi^{2}+gf+\beta=0
\end{equation}
or in terms of $y^{2}=f$
\begin{equation}\label{oss}
y^{\prime\prime}+\frac{\gamma}{\alpha}\xi^{2}y+\frac{g}{\alpha}y^{3}+\beta
y=0.
\end{equation}
The equation for the dispersion $d$ can be integrated once to
\begin{equation}
(d^{\prime})^{2}=\epsilon d^{2}+4\alpha^{3}\gamma.\label{tre}
\end{equation}
If $\alpha\gamma>0,$ $\epsilon<0$ we have periodical solutions for $d(t),$ but
the equation for $f$ has no finite solutions; if $\alpha\gamma<0,\,\epsilon
>0,$ one can obtain the solution
\begin{equation}
d=\sqrt{-\frac{4\alpha^{3}\gamma}{\epsilon}}\,\cosh\,(t\sqrt{\epsilon
}).\label{toch}
\end{equation}
We say we have obtain an exact self-similar solution of NLS -
equation with time dependent dispersion because solutions of
(\ref{oss}) have no divergences and have well-defined asymptotical
behavior (Schr\~odinger equation for oscillator with cubic
nonlinearity). It is important that we have no exact solutions in
a case of periodic dispersion. So, an optical soliton
\cite{has, tur} in the periodic
case may be non-stable.

We have obtain the self-similar equations for NLS-type dynamical
systems. It is possible to apply the Lagrangian method to the integrable
two-dimensional systems or to non-integrable systems.
\section*{Appendices}
\appendix
\section{Miura transformations for self-similar solutions}

The sequence of Miura transformations (\ref{mi}) leads to analogous ones for
the self-similar solutions:
\begin{equation}
\text{isotropic Heisenberg\ model\thinspace(I)}\rightarrow
\text{Volterra\ model\thinspace(II)}\rightarrow\text{Toda\ model\thinspace
(III)}\label{mis}
\end{equation}
\[
I\rightarrow II:\quad-zy+\frac{z^{\prime}}{z}\rightarrow z,\quad-zy\rightarrow
y,\quad2(1-\nu)=c_{2}-c_{1},\quad\mu=c_{1}
\]
\[
II\rightarrow III:\quad zy+z^{\prime}\rightarrow z,\quad z+y\rightarrow
y,\quad C=c_{1}+c_{2},\quad D=c_{2}(c_{1}+2).
\]
For obtaining the relation among constants of different reduced systems, in
the case $I\rightarrow II,$ one inserts the Miura transformed solutions $z$
and $y$ into the system $II$ getting for the first equation the conservation
law (\ref{mu}) containing the constant $c_{1}$ and for the second equation the
same conservation law containing the constant $c_{2},$ while, in the case
$II\rightarrow III,$ one inserts the Miura transformed solutions $z$ and $y$
into the conservation law (\ref{Todaconservation}) of system $III$ and then
uses the second equation of the system $II$ and the PIV equation for the
self-similar solution $y$ of the system $II.$

It is interesting to note that the Miura transformation $II\rightarrow III$ by
using the first equation of the system $II$ can be inverted getting
\[
III\rightarrow II:\quad y-\frac{y^{\prime}+c_{1}-z}{y-2\xi}\rightarrow
z,\quad\frac{y^{\prime}+c_{1}-z}{y-2\xi}\rightarrow y.
\]
\section{Lagrangians for Painlev\'{e} equations}
The six Painlev\'{e} equations ($a,b,c,d$ constants)
\begin{align}
P1:\qquad w_{xx} &  =6w^{2}+x,\label{p1}\\
P2:\qquad w_{xx} &  =2w^{3}+xw+a,\label{p2}\\
P3:\qquad w_{xx} &  =\frac{w_{x}^{2}}{w}-\frac{w_{x}}{x}+\frac{aw^{2}+b}
{x}+cw^{3}+\frac{d}{w},\label{p3}\\
P4:\qquad w_{xx} &  =\frac{w_{x}^{2}}{2w}+\frac{3w^{3}}{2}+4xw^{2}
+2(x^{2}-a)w+\frac{b}{w},\label{p4}\\
P5:\qquad w_{xx} &  =(\frac{1}{2w}+\frac{1}{w-1})w_{x}^{2}-\frac{w_{x}}
{x}+\frac{2}{x^{2}}(w-1)^{2}(aw+\frac{b}{w})+\frac{cw}{x}+\frac{dw(w+1)}
{w-1},\label{p5}\\
P6:\qquad w_{xx} &  =\frac{1}{2}(\frac{1}{w}+\frac{1}{w-1}+\frac{1}{w-x}
)w_{x}^{2}-(\frac{1}{x}+\frac{1}{x-1}+\frac{1}{w-x})w_{x}+\label{p6}\\
&  \quad\quad+\frac{w(w-1)(w-x)}{x^{2}(x-1)^{2}}\left[  a+\frac{bx}{w^{2}
}+\frac{c}{(w-1)^{2}}(x-1)+\frac{dx(x-1)}{(w-x)^{2}}\right]  ,\nonumber
\end{align}
can be derived by variation of the following Lagrangians:
\begin{align}
L1 &  :\qquad L=\frac{w_{x}^{2}}{2}+2w^{3}+xw,\label{l1}\\
L2 &  :\qquad L=w_{x}^{2}+w^{4}+x\,w^{2}+2aw,\label{l2}\\
L3 &  :\qquad L=\frac{w_{x}^{2}}{2w^{2}}x+aw-\frac{b}{w}+\frac{x}{2}
(cw^{2}-\frac{d}{w^{2}}),\label{l3}\\
L4 &  :\qquad L=\frac{w_{x}^{2}}{w}+w^{3}+4w^{2}x+4w(x^{2}-a)-\frac{2b}
{w},\label{l4}\\
L5 &  :\qquad L=\frac{w_{x}^{2}}{w(w-1)^{2}}x+\frac{4}{x}(aw-\frac{b}
{w})-\frac{2c}{w-1}-\frac{2dw}{(w-1)^{2}}x,\label{l5}\\
L6 &  :\qquad L=\frac{w_{x}^{2}}{2}\frac{x(x-1)}{w(w-1)(w-x)}+\frac
{aw}{x(x-1)}-\frac{b}{w(x-1)}-\frac{c}{x(w-1)}-\frac{d}{w-x}.\label{l6}
\end{align}

\end{document}